\documentclass[a4paper, 12pt]{article}

\usepackage[sort&compress]{natbib}
\bibpunct{(}{)}{;}{a}{}{,} 
\usepackage{amsthm, amsmath, amssymb, mathrsfs, multirow, url}
\usepackage{graphicx} 
\usepackage{ifthen} 
\usepackage{amsfonts}
\usepackage[usenames]{color}
\usepackage{fullpage}
\usepackage{subfigure}

\theoremstyle{plain}

\theoremstyle{definition}

\theoremstyle{remark} 
\newtheorem*{pralg}{Predictive Recursion Algorithm}

\newcommand{\E}{\mathsf{E}}
\newcommand{\V}{\mathsf{V}}

\newcommand{\unif}{{\sf Unif}}
\newcommand{\nm}{{\sf N}}

\newcommand{\gam}{{\sf Gamma}}
\newcommand{\stt}{{\sf t}}

\newcommand{\bet}{{\sf Beta}}

\newcommand{\RR}{\mathbb{R}}

\newcommand{\XX}{\mathbb{X}}
\newcommand{\YY}{\mathbb{Y}}

\renewcommand{\S}{\mathcal{S}}

\title{Permutation-based uncertainty quantification about a mixing distribution}
\author{Vaidehi Dixit\footnote{Department of Statistics, North Carolina State University; {\tt vdixit@ncsu.edu}, {\tt rgmarti3@ncsu.edu}} \quad and \quad Ryan Martin$^*$}
\date{\today}

\begin{document}

\maketitle 

\begin{abstract}    
Nonparametric estimation of a mixing distribution based on data coming from a mixture model is a challenging problem.  Beyond estimation, there is interest in uncertainty quantification, e.g., confidence intervals for features of the mixing distribution.  This paper focuses on estimation via the predictive recursion algorithm, and here we take advantage of this estimator's seemingly undesirable dependence on the data ordering to obtain a permutation-based approximation of the sampling distribution which can be used to quantify uncertainty. Theoretical and numerical results confirm that the proposed method leads to valid confidence intervals, at least approximately.  

\smallskip

\end{abstract}

\section{Introduction}
\label{S:intro}  

At a high-level, statistical analysis aims to separate signal from noise, and one of the more challenging problems is deconvolution or, more generally, estimation of a mixing distribution based on samples from the mixture.  Suppose that data $Y^n = (Y_1,\ldots,Y_n)$ are independent and identically distributed from a density $f$ with respect to Lebesgue measure on $\YY$, which we model as a mixture 
\begin{equation}
\label{eq:mixture}
f(y) = \int_\XX k(y \mid x) \, p(x) \, \mu(dx), \quad y \in \YY, 
\end{equation}
where $k$ is a known kernel, i.e., $y \mapsto k(y \mid x)$ is a density on $\YY$ for each $x \in \XX$, and $p$ is a unknown mixing density with respect to a known $\sigma$-finite measure $\mu$ on $\XX$.  Alternatively, one can view this model hierarchically by assuming that $X_1,\ldots,X_n$ are independent and identically distributed from $p$, and $Y_i$, given $X_i$, are independently distributed from $k(y \mid X_i)$, $i=1,\ldots,n$.  So if we think of $X_1,\ldots,X_n$ as ``signals'' with distribution $p$, and $Y_i$ the version corrupted by noise, then our goal---inference about $p$ based on data from model \eqref{eq:mixture}---can be viewed as separation of signal from noise.  

Often, $p$ is assumed to be discrete with finitely many points in $\XX$.  For these so-called finite mixture models, standard modes of inference can be applied.  For example, when the number of support points of $p$ is known, there is a likelihood function with relatively simple form that can be optimized, often using the EM algorithm \citep{dempetal, teeletal}, to produce the corresponding maximum likelihood estimator, to which the classical asymptotic distribution theory applies \citep[e.g.,][]{rednerwalker}; for a comprehensive treatment, see \citet{mcpeel}.  Similarly, with this same likelihood and a corresponding prior distribution for $p$, one can apply an EM-like data-augmentation strategy \citep[e.g.,][]{vm} to carry out a Bayesian analysis.  In the more realistic scenario where the number of components in the finite mixture model is unknown, those methods described above can be modified by introducing a penalty term or a prior distribution on the number of mixture components, as in \citet{leroux} and \citet{richardgreen}.  


For the case considered here, where $p$ is a smooth mixing density, a number of methods for nonparametric estimation have appeared in the literature.  When $k(y \mid x) = k(y - x)$ is a location-shift kernel, so that the mixture density is just a convolution, estimation of $p$ is referred to as deconvolution, a case that has been studied in \citet{fan}, \citet{sc}, and \citet{zhang}.  For general mixtures, there are a variety of different approaches.  Maximizing the likelihood will almost surely produce a discrete estimate of the mixing distribution \citep{lindsay}, which is not a satisfactory estimate of a smooth mixing density.  Various approaches aim to smooth the discrete nonparametric maximum likelihood estimator, either directly \citep[e.g.,][]{eggermont} or by introducing a smoothness penalty \citep[e.g.,][]{liuetall}.  \citet{chaeetall} investigate an iterative algorithm that generates a sequence of smooth mixing density estimates that converge to the nonparametric maximum likelihood estimator.  Another interesting and related method, which is the focus of the present paper, is that based on the {\em predictive recursion} algorithm first described in \citet{nqz}, \citet{newtonzhang}, and \citet{newton02}, with extensions and theoretical properties developed in \citet{ghoshtokdar}, \citet{martinghosh}, \citet{tmg}, and \citet{mt-rate, mt-prml, mt-test}; for a recent review of these developments, see \citet{pr.jkg.review}.

Beyond estimation, a goal is to quantify uncertainty about the mixing density $p$ and, for this, the literature is scarce. The work that has been done is as listed below. \citet{bd} discuss asymptotic and bootstrap confidence bands for deconvolution problems, building on ideas first presented in \citet{br}. \citet{ls} discuss empirical Bayes estimation of an empirical mixing distribution with emphasis on construction of interval estimates, using the nonparametric maximum likelihood estimator.  \citet{fortpetrone} have developed asymptotically approximate credible intervals for the cumulative distribution function based on a quasi-Bayesian interpretation of the predictive recursion algorithm.  


A seemingly undesirable feature of the predictive recursion estimator is that it depends on the order in which the data is processed.  In particular, this means that the estimator is not a function of the sufficient statistic---which, in this setting, is the empirical distribution---and, hence, the estimator cannot be Bayesian.  In previous literature on predictive recursion, the focus has been on reducing its dependence on the order.  For example, \citet{newton02} suggested elimination of the order-dependence by averaging the estimators over a number of randomly chosen permutations; see \citet{tmg} for details.  The idea in the present paper is to leverage predictive recursion's order-dependence for the purpose of uncertainty quantification.  Specifically, we propose to generate multiple copies of the predictive recursion estimator by permuting the data sequence, and then use this permutation-based distribution as an approximation of the estimator's sampling distribution.  After a review of the predictive recursion estimator and its properties in Section~\ref{S:pr}, we show in Section~\ref{S:perm} that, for a given feature of the mixing distribution, the estimator's sampling distribution variance can be approximately unbiasedly estimated by the corresponding permutation-based distribution variance.  Being able to accurately estimate the spread of the relevant sampling distribution immediately suggests using the same permutation distribution quantiles as an approximate confidence interval for that mixing distribution feature.  Numerical results presented in Section~\ref{S:examples} reveal that this permutation-based approach gives approximately valid confidence intervals in finite samples across a range of mixture models and for various features of the mixing distribution, including the density function at a point.

\section{Predictive recursion}
\label{S:pr}

Recall that we observe independent data $Y^n = (Y_1,\ldots,Y_n)$ from the mixture in \eqref{eq:mixture}, and the goal is to estimate the mixing density $p$.  The following predictive recursion algorithm returns a computationally efficient nonparametric estimator of $p$.  

\begin{pralg}
Start with an initial estimate $p_0$ of the mixing density and a sequence of weights $\{w_i : i \geq 1\} \subset (0,1)$. Using the observations $Y_1,\ldots,Y_n$ from the mixture model, in that order, compute
\begin{equation}
\label{eq:pr}
p_i(x) = (1-w_i) \, p_{i-1}(x) + w_i \, \frac{k(Y_i \mid x)p_{i-1}(x)}{f_{i-1}(Y_i)}, \quad i=1,\ldots,n,
\end{equation}
where $f_{i-1}(y) = \int k(y \mid x)p_{i-1}(x) \, \mu(dx)$ is the mixture corresponding to $p_{i-1}$.  Finally, return $p_n$ and $f_n=f_{p_n}$ as the estimates of $p$ and $f$, respectively.
\end{pralg}

An interesting observation is that, if $p_0$ is a smooth density with respect to the measure $\mu$, then the output, $p_n$, of the predictive recursion algorithm will also be a smooth density.  Compare this to the nonparametric MLE which is almost surely discrete, regardless of the smoothness of $p$ in \eqref{eq:mixture}.  Therefore, there is no need for post hoc smoothing of the predictive recursion estimator.  And the ability to specify the dominating measure in the predictive recursion algorithm proved to be a useful property in the multiple testing application considered in \citet{mt-test}.  

Aside from estimating the mixing density itself, one can readily estimate various features of the mixing distribution.  That is, if $\psi$ is a suitable function, then $\int \psi \, p \, d\mu$ can be estimated by $\int \psi \, p_n \, d\mu$.  For example, we can estimate the mixing distribution function at a point $x_0$ by taking $\psi$ to be the indicator function corresponding to $(-\infty,x_0]$.  

Asymptotic convergence properties of the predictive recursion estimator were investigated in \citet{tmg} and \citet{mt-rate}.  To summarize, under suitable tail conditions on the kernel, if $p$ is identifiable from the mixture model \eqref{eq:mixture} and if the weights $(w_i)$ in the predictive recursion algorithm satisfy 
\[ \sum_{i=1}^\infty w_i = \infty \quad \text{and} \quad \sum_{i=1}^\infty w_i^2 < \infty, \]
then $p_n$ converges to $p$ almost surely in the weak topology, that is, if $\psi: \XX \to \RR$ is a bounded and continuous function, then $\int \psi \, p_n \,d\mu \to \int \psi \, p \, d\mu$ almost surely.

\section{Leveraging order-dependence}
\label{S:perm}

It is clear from \eqref{eq:pr} that $p_n$ depends on the order in which the data are processed.  However, since the data are assumed to be independent and identically distributed, the ordering should be irrelevant.  To alleviate predictive recursion's seemingly undesirable order-dependence, \citet{newton02} and others have  suggested to average the final estimate, $p_n$, over a number of randomly chosen permutations of the data sequence.  \citet{tmg} describe this as a sort of Rao--Blackwellization, replacing $p_n$ by a Monte Carlo approximation of its conditional expectation given the order statistics.  Here, instead of trying to remove predictive recursion's order-dependence, we propose to leverage it for the purpose of uncertainty quantification.  

Let $\S_n$ denote the permutation group on integers $\{1,2,\ldots,n\}$, i.e., the set of all bijections from $\{1,2,\ldots,n\}$ to itself.  If $p_n$ is the predictive recursion estimator based on data $(Y_1,\ldots,Y_n)$, in its given order, write $p_n^s$ for the corresponding estimator based on the data $Y_{s(1)},\ldots,Y_{s(n)}$, permuted according to $s \in \S_n$.  If $\psi$ is a suitable function, write $\Psi_n = \int \psi \, p_n \, d\mu$ and $\Psi_n^s = \int \psi \, p_n^s \,d\mu$ as the estimators of $\Psi = \int \psi \, p \, d\mu$ based on the original and permuted data, respectively.  

Our proposal is to approximate the sampling distribution of $\Psi_n$, as a function of $Y^n$ sampled from the mixture in \eqref{eq:mixture}, by the distribution of $\Psi_n^S$, for fixed $Y^n$, as a function of $S \sim \unif(\S_n)$.  The justification for our claim that this provides an accurate approximation, at least asymptotically, is the following simple identity, 
\begin{equation}
\label{eq:identity}
\V_{Y^n}( \Psi_n ) = \V_{Y^n,S}( \Psi_n^S ), 
\end{equation}
where $\V_{Y^n}$ is the variance with respect to the distribution of $Y^n$ from the mixture model \eqref{eq:mixture}, and $\V_{Y^n,S}$ is the variance with respect to the joint distribution of $Y^n$ and $S \sim \unif(\S_n)$, assumed to be independent.  The identity \eqref{eq:identity} holds because, when data $Y^n$ are independent and identically distributed, the extra layer of permutations changes nothing.  In other words, if we imagine enumerating all possible realizations of $Y^n$ to evaluate the left-hand side of \eqref{eq:identity}, then we would get no new realizations if we also enumerated permutations for evaluating the right-hand side.  

The practical value of the identity \eqref{eq:identity} is that the right-hand side suggests a familiar total-variance decomposition:
\begin{equation}
\label{eq:decomp}
\V_{Y^n, S}(\Psi_n^S) = \E_{Y^n}\{ \V_{Y^n,S}( \Psi_n^S \mid Y^n)\} + \V_{Y^n}\{ \E_{Y^n, S}( \Psi_n^S \mid Y^n) \}. 
\end{equation}
By the asymptotic consistency property of predictive recursion, discussed in Section~\ref{S:pr}, if $\Psi$ is bounded and continuous, then $\Psi_n$ and, hence, $\E_{Y^n,S}(\Psi_n^S \mid Y^n)$ are consistent estimates of $\Psi$, so its variance, the second term on the right-hand side of \eqref{eq:decomp}, should be near 0 when $n$ is large.  Therefore, 
\begin{equation}
\label{eq:unbiased}
\V_{Y^n, S}(\Psi_n^S) \approx \E_{Y^n}\{ \V_{Y^n,S}( \Psi_n^S \mid Y^n)\}, 
\end{equation}
in other words, $\V_{Y^n,S}( \Psi_n^S \mid Y^n)$ is an approximately unbiased estimator of $\V_{Y^n}(\Psi_n)$.  The key point, of course, is that the left-hand side of \eqref{eq:identity}, namely, $\V_{Y^n}(\Psi_n)$, the variance of the sampling distribution of $\Psi_n$, is relevant for uncertainty quantification, but it is not readily available.  However, the quantity on the right-hand side of \eqref{eq:identity}, namely, $\V_{Y^n,S}(\Psi_n^S \mid Y^n)$, can be readily computed by repeatedly sampling $S \sim \unif(\S_n)$, permuting the data according to $S$, and re-evaluating the predictive recursion estimator.   This provides us with a relatively simple and fast approach to construct a data-dependent distribution for $p$ or $\Psi$ from which valid uncertainty quantification can be achieved.  And beyond variance estimation, to get an approximate $100(1-\alpha)$\% confidence interval for $\Psi$, we can easily extract the $\frac{\alpha}{2}$ and $1-\frac{\alpha}{2}$ quantiles of $\Psi_n^S$ based on repeated sampling of $S \sim \unif(\S_n)$ with data $Y^n$ fixed.

\section{Numerical results}
\label{S:examples}

Here we carry out an empirical investigation into the performance of our proposed permutation-based approach to uncertainty quantification.  We begin with pointwise evaluation of the mixing distribution function.  The specific scenario we consider here is one where the true mixing density in \eqref{eq:mixture} is a gamma with shape 2 and rate 1, and the kernel $k(y \mid x)$ is also a gamma with shape $20x$ and rate 20; this is Example~3-3 below.  We generate samples of size $n=500$ from this mixture model and consider estimating the mixing distribution function, in particular, at the fixed points $x \in \{2, 5, 8\}$.  For predictive recursion, we take  initial guess $p_0 = \unif(0,10)$ and weight sequence $w_i = (i+1)^{-0.67}$ suggested by \citet{mt-rate}.

Before presenting the results, we have to address a relevant and practically important question, namely, how many permutations?  In our examples, we are using 200 randomly generated permutations on which to evaluate the predictive recursion estimator.  Of course, it does not hurt to do more than 200, and this is still computationally feasible thanks to the algorithm's efficiency; however, at least in the examples we tried, there were no substantial differences in the results based on more than 200 random permutations.  

\begin{figure}[t]
\begin{center}
\scalebox{0.7}{\includegraphics{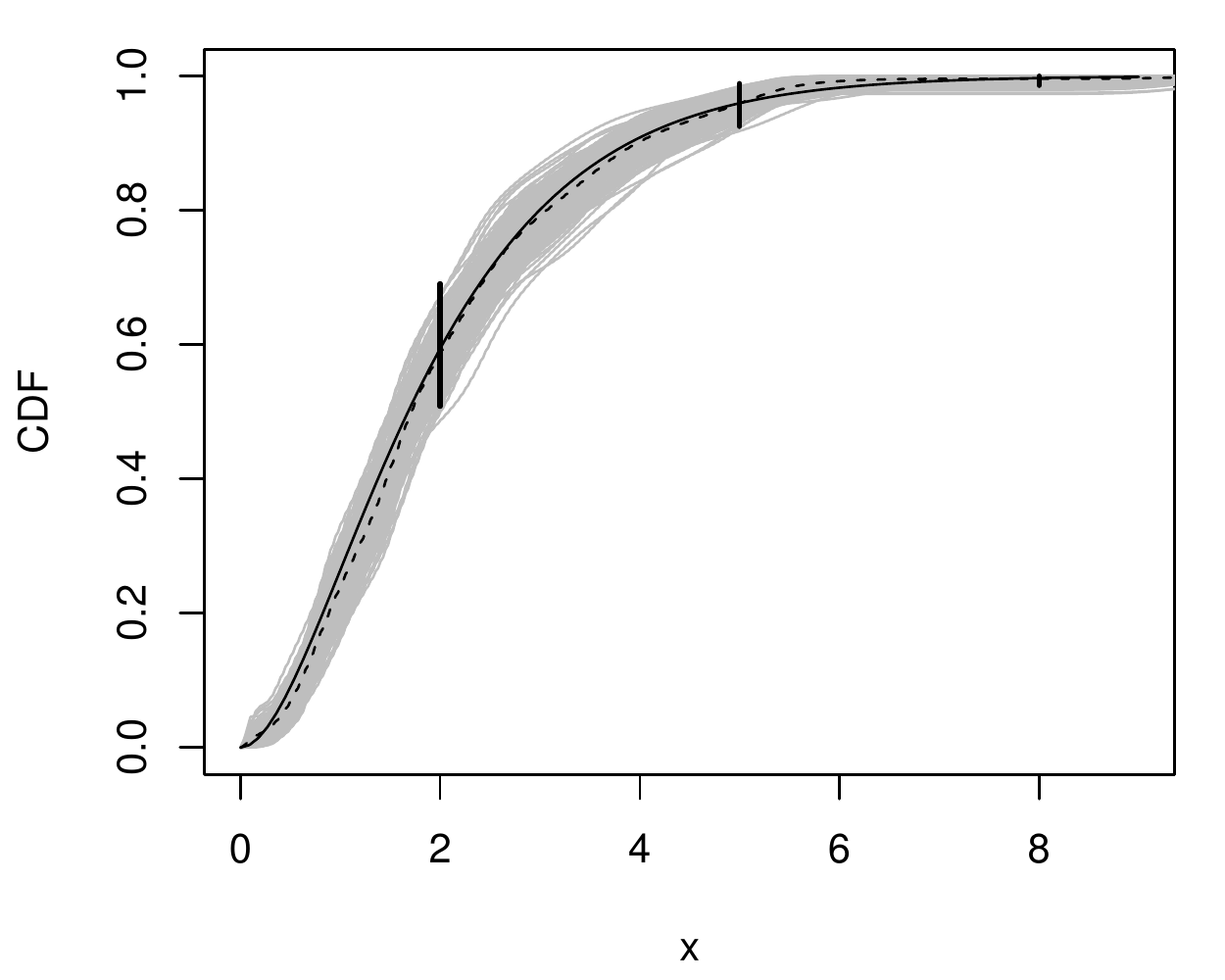}}
\end{center}
\caption{Plot of the predictive recursion estimates of the distribution function for each of 200 random permutations (gray), along with the true distribution function (black); dashed line corresponds to the predictive recursion estimate averaged over permutations. Vertical bars correspond to the central 95\% interval of the sampling distribution of the predictive recursion estimator.}
\label{fig:cdf}
\end{figure}

A plot of distribution function estimates for this gamma mixture example, based on 200 random permutations, is shown in Figure~\ref{fig:cdf}.  Note that the span of these estimates over permutations hugs the true distribution relatively closely across the entire range of $x$.  The vertical bars at $x \in \{2, 5, 8\}$ correspond to the central 95\% interval of the sampling distribution of the predictive recursion estimator, based on repeated sampling from the gamma mixture.  The goal of uncertainty quantification is to match this interval as closely as possible, so it is notable that, as predicted by the arguments leading up to \eqref{eq:unbiased}, our permutation-based intervals are comparable to this ``gold-standard'' across various $x$.  Moreover, the permutation distribution takes only about 10 seconds in R running on an ordinary laptop computer.  \citet{fortpetrone} present asymptotically approximate credible intervals for the same cumulative distribution function based on the predictive recursion algorithm.  But their analysis is based on a dependent and non-stationary model for $Y^n$---one that makes the predictive recursion estimator ``quasi-Bayes''---and, since their model and perspective on uncertainty quantification is different from ours, a direct comparison is not appropriate.  

Next we consider pointwise estimation of the mixing density.  This is not strictly covered by the theoretical arguments discussed above because the functional $p \mapsto p(x)$ for a fixed $x$ cannot be expressed as $\int \psi \, p \, d\mu$ for a bounded and continuous $\psi$.  However, intuition and prior experience suggests that the predictive recursion density estimate ought to satisfy a pointwise consistency property, i.e., $p_n(x) \to p(x)$ for fixed $x$; see, also, Section~\ref{S:discuss}.  Therefore, we can apply the same total-variance decomposition as in \eqref{eq:decomp} and reason that 
\[ \V_{Y^n}\{p_n(x)\} \approx \E_{Y^n}\bigl[ \V_{Y^n,S}\{ p_n^S(x) \mid Y^n\} \bigr] \]
and, moreover, that suitable quantiles from the permutation distribution can be used in the obvious way to construct approximate confidence intervals for $p(x)$.  

Like in \citet{chaeetall}, we consider nine examples corresponding to different combinations of the following three kernels and mixing densities.  
\begin{description}
\item[\sc Kernel 1.] $k(y \mid x) = \nm(y \mid x, 0.5)$;
\item[\sc Kernel 2.] $k(y \mid x) = \frac{1}{0.3}\stt(\frac{y-x}{0.3} \mid \text{df}=5)$;
\item[\sc Kernel 3.] $k(y \mid x) = \gam(x \mid \text{shape}=20x,\, \text{rate}=20)$;
\item[\sc Mixing Density 1.] $p(x) =  \frac{1}{10}\bet(\frac{x}{10} \mid 5, 5)$;
\item[\sc Mixing Density 2.] $p(x) = \frac{3}{4}\nm(x \mid 3,0.8^{2}) + \frac{1}{4}\nm(x \mid 7,0.8^{2})$;
\item[\sc Mixing Density 3.] $p(x) = \gam(x \mid \text{shape}=2, \, \text{rate}=1)$.
\end{description}
In what follows, Example~$a$-$b$ will refer to the case with kernel $a$ and mixing density $b$, where $a=1,2,3$ and $b=1,2,3$.  

\begin{figure}[t]
\begin{center}
\subfigure[Example 1-1]{\includegraphics[width = 0.3\linewidth]{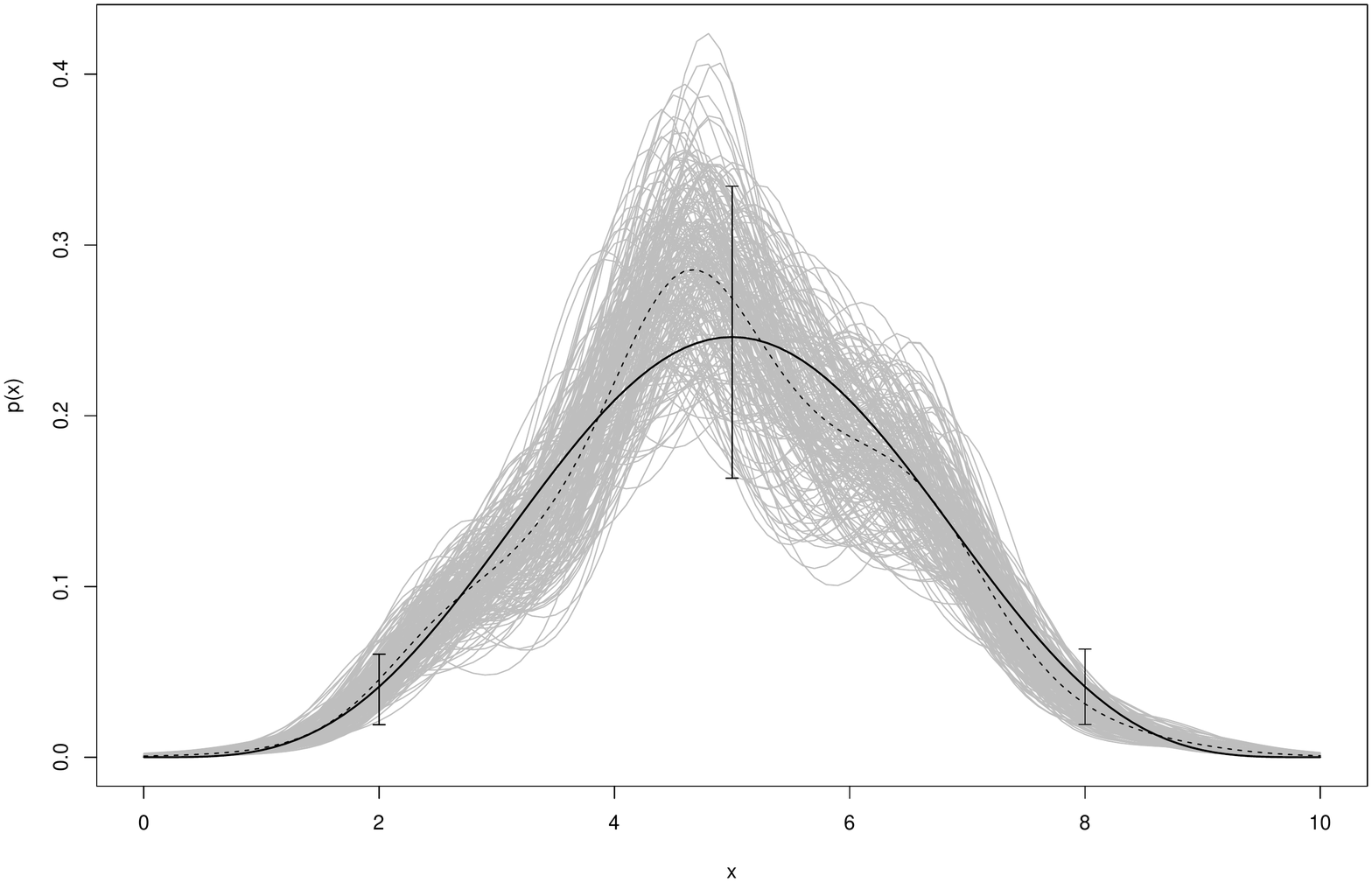}} 
\subfigure[Example 2-1]{\includegraphics[width = 0.3\linewidth]{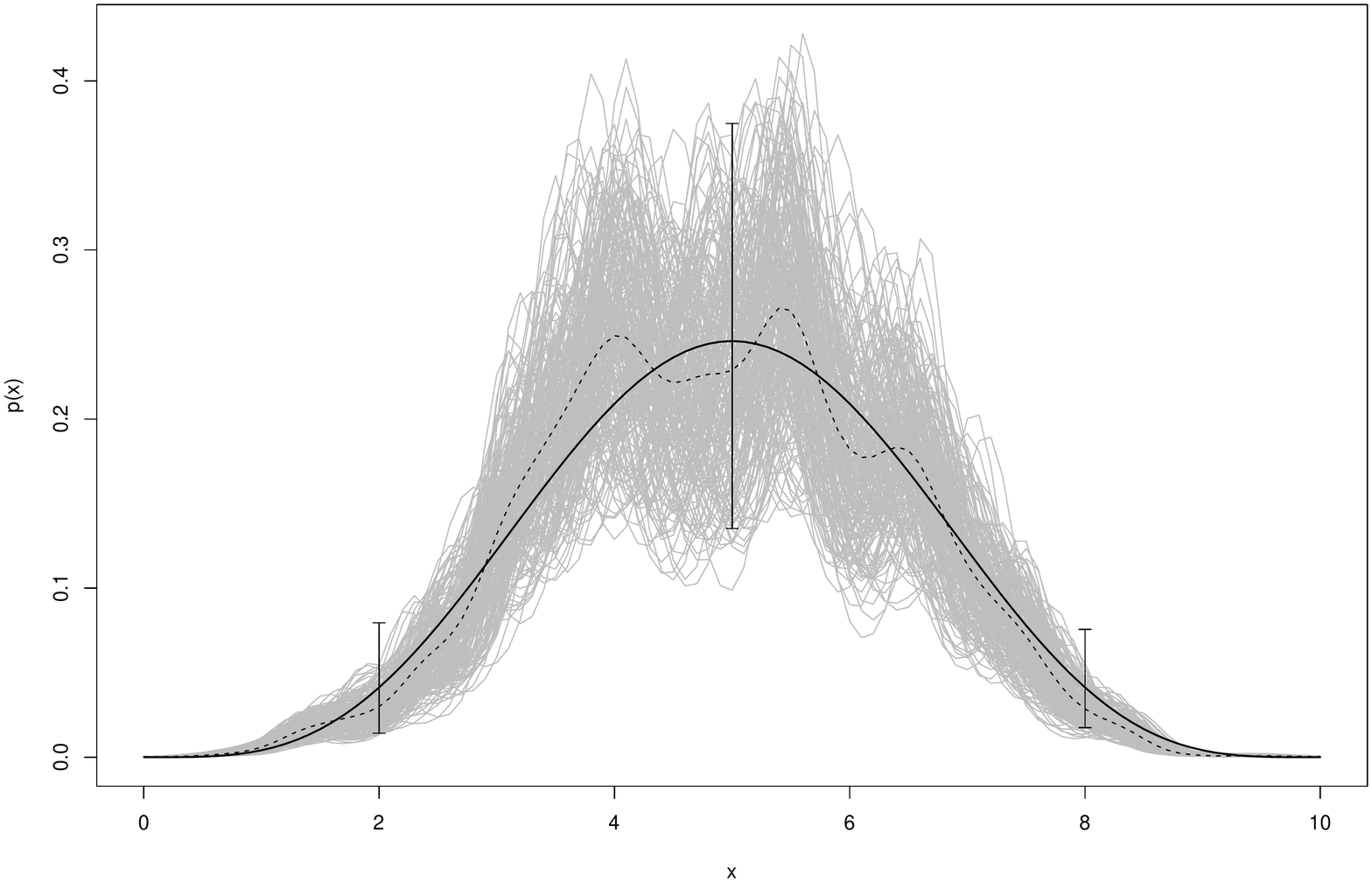}} 
\subfigure[Example 3-1]{\includegraphics[width = 0.3\linewidth]{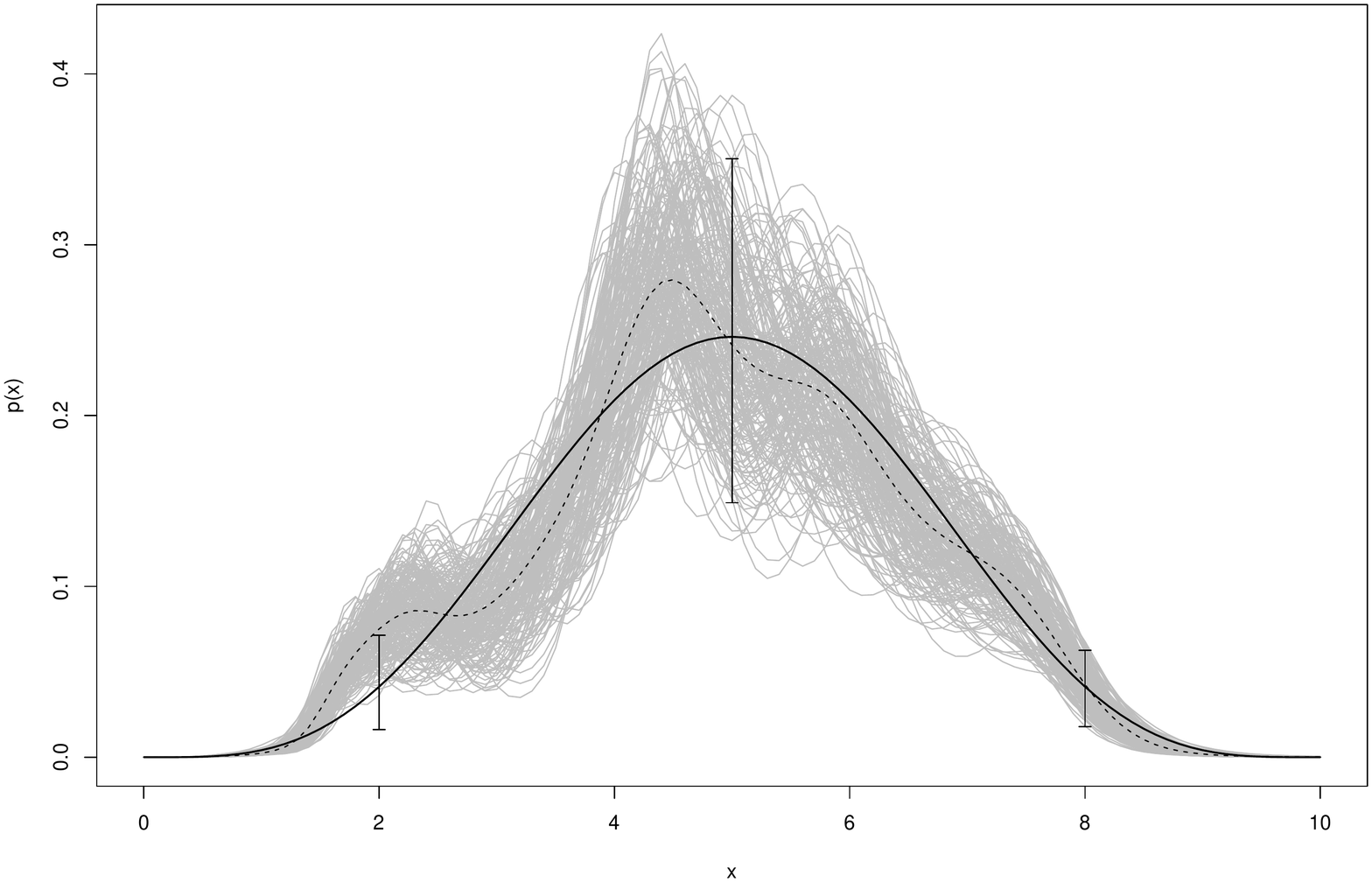}} 
\subfigure[Example 1-2]{\includegraphics[width = 0.3\linewidth]{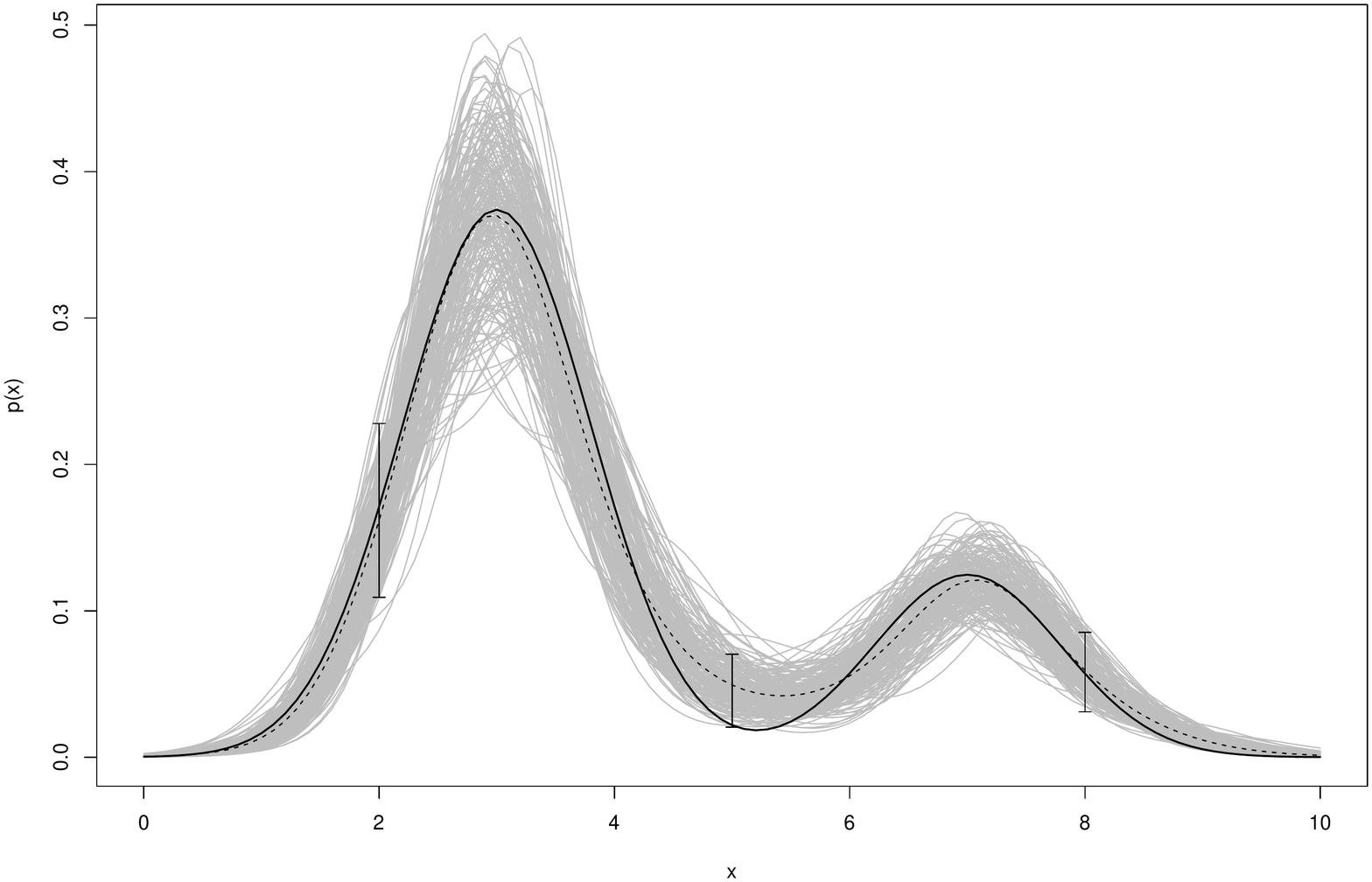}}
\subfigure[Example 2-2]{\includegraphics[width = 0.3\linewidth]{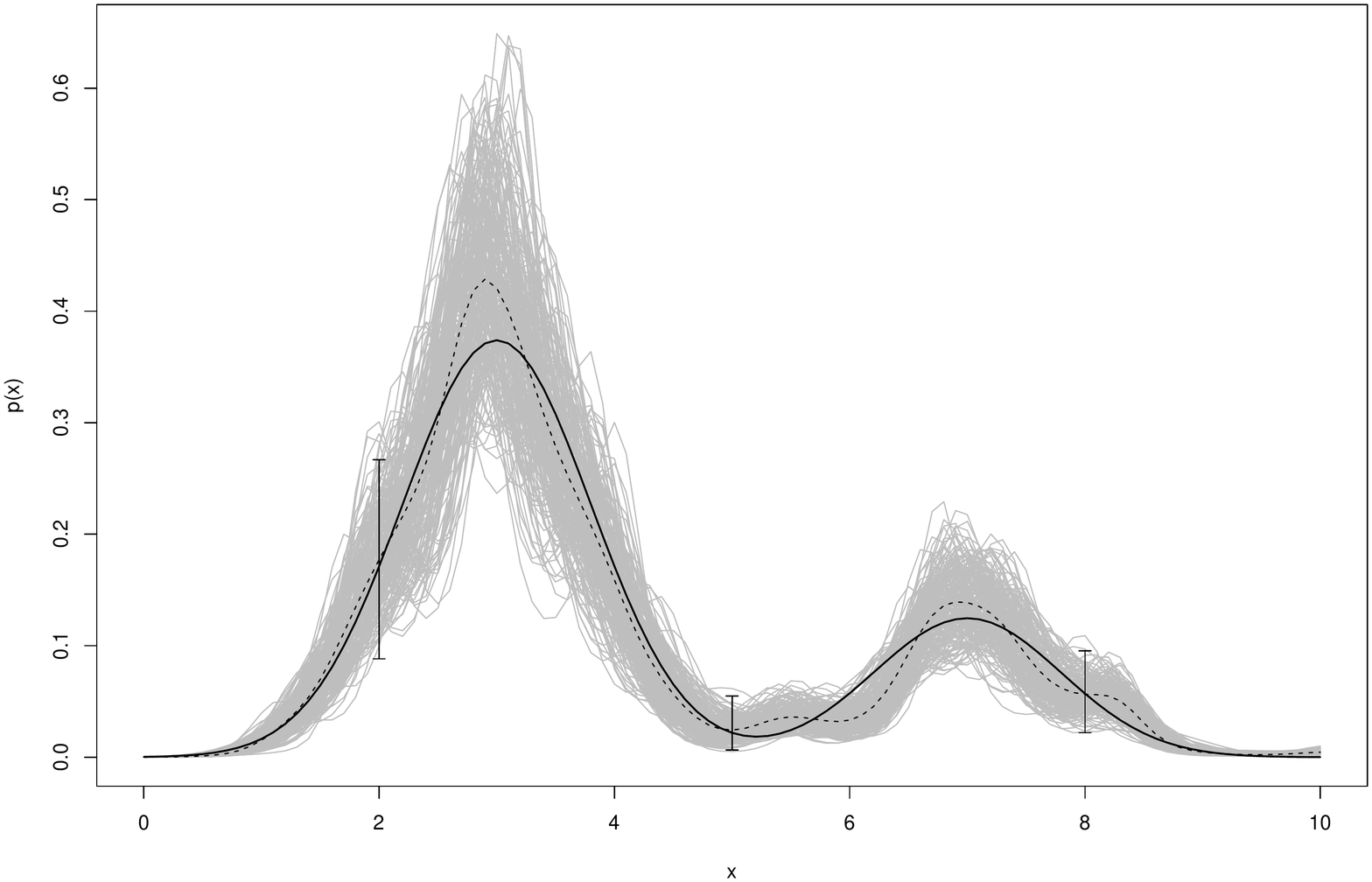}} 
\subfigure[Example 3-2]{\includegraphics[width = 0.3\linewidth]{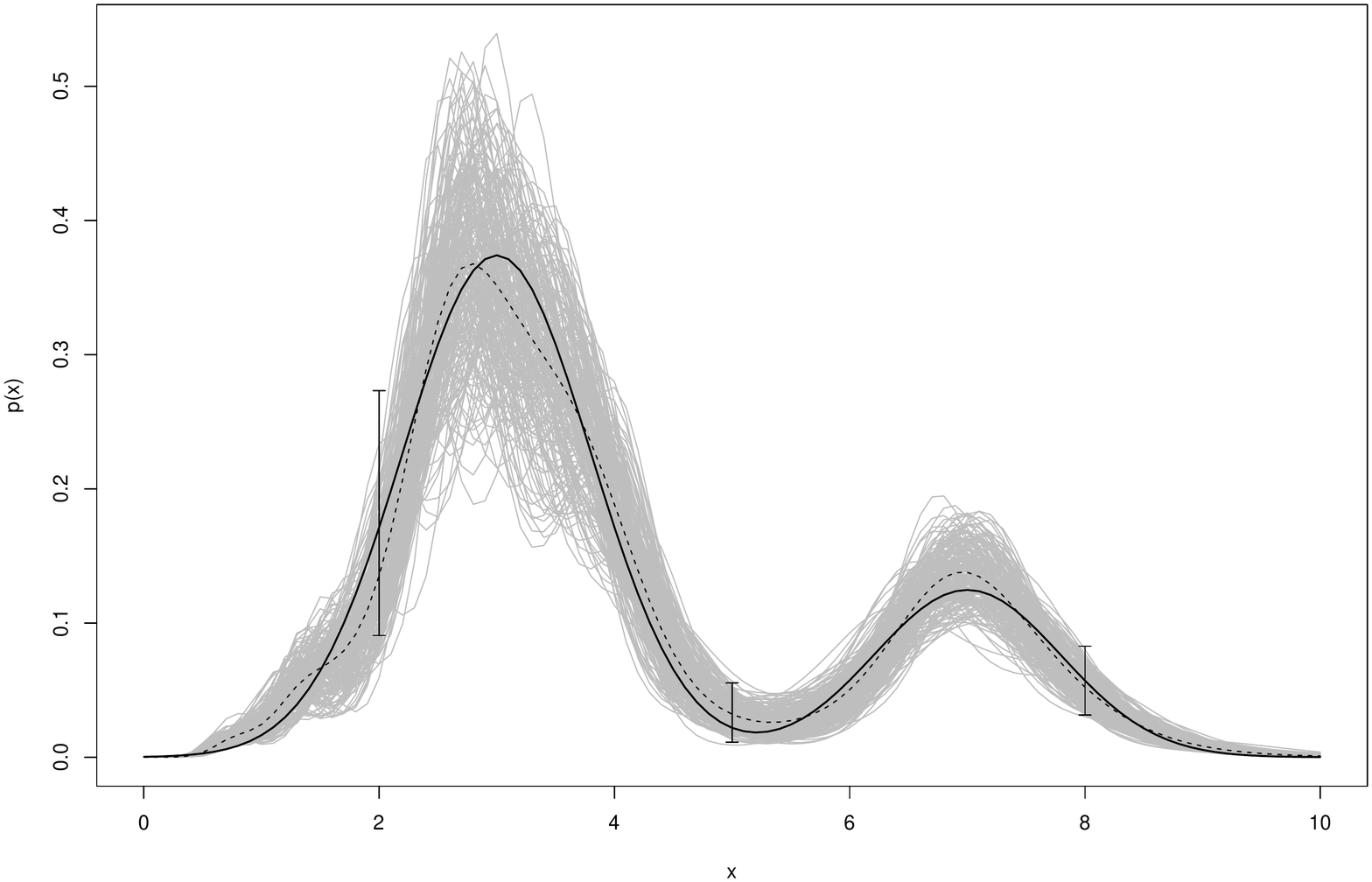}} 
\subfigure[Example 1-3]{\includegraphics[width = 0.3\linewidth]{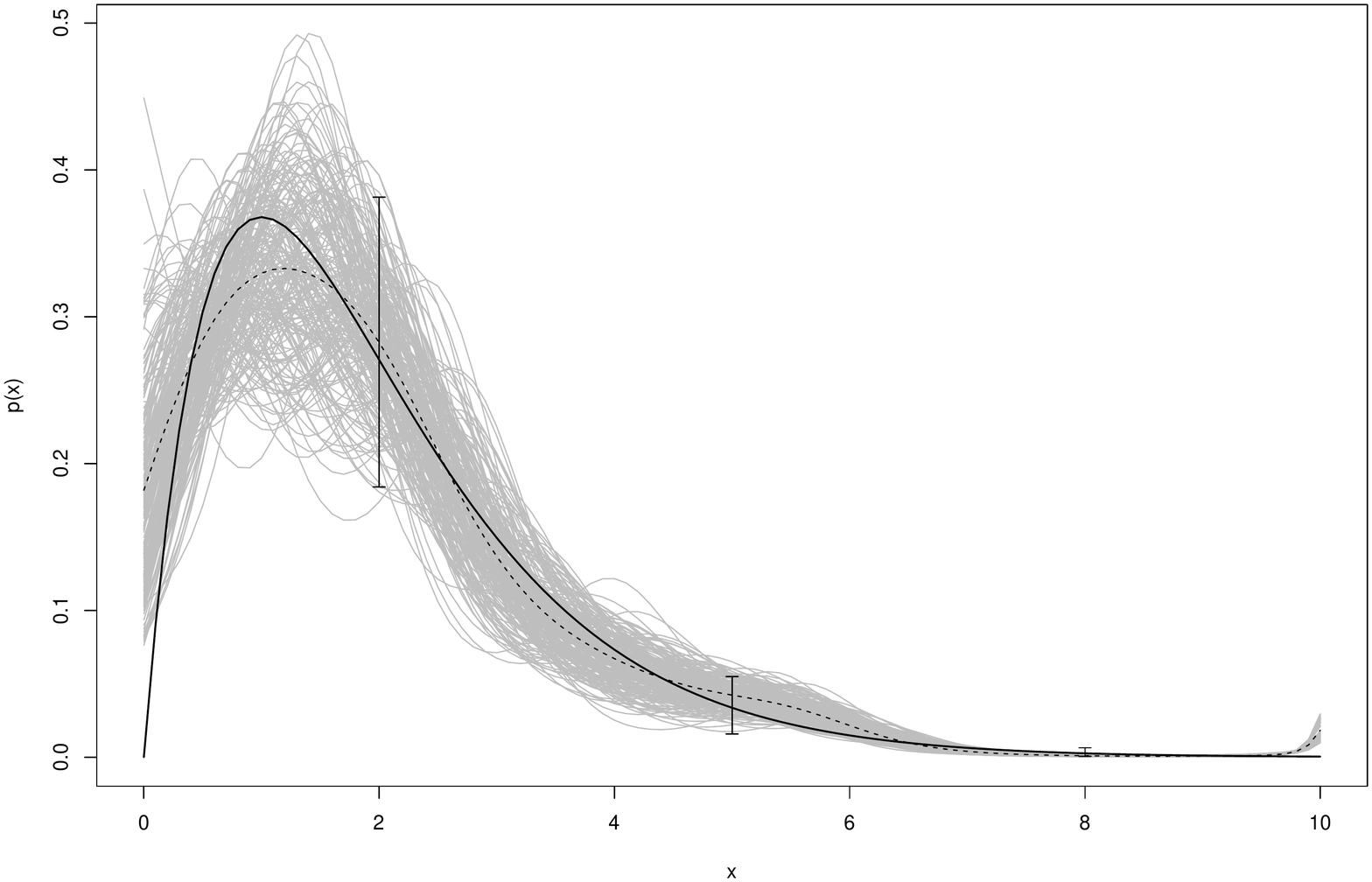}} 
\subfigure[Example 2-3]{\includegraphics[width = 0.3\linewidth]{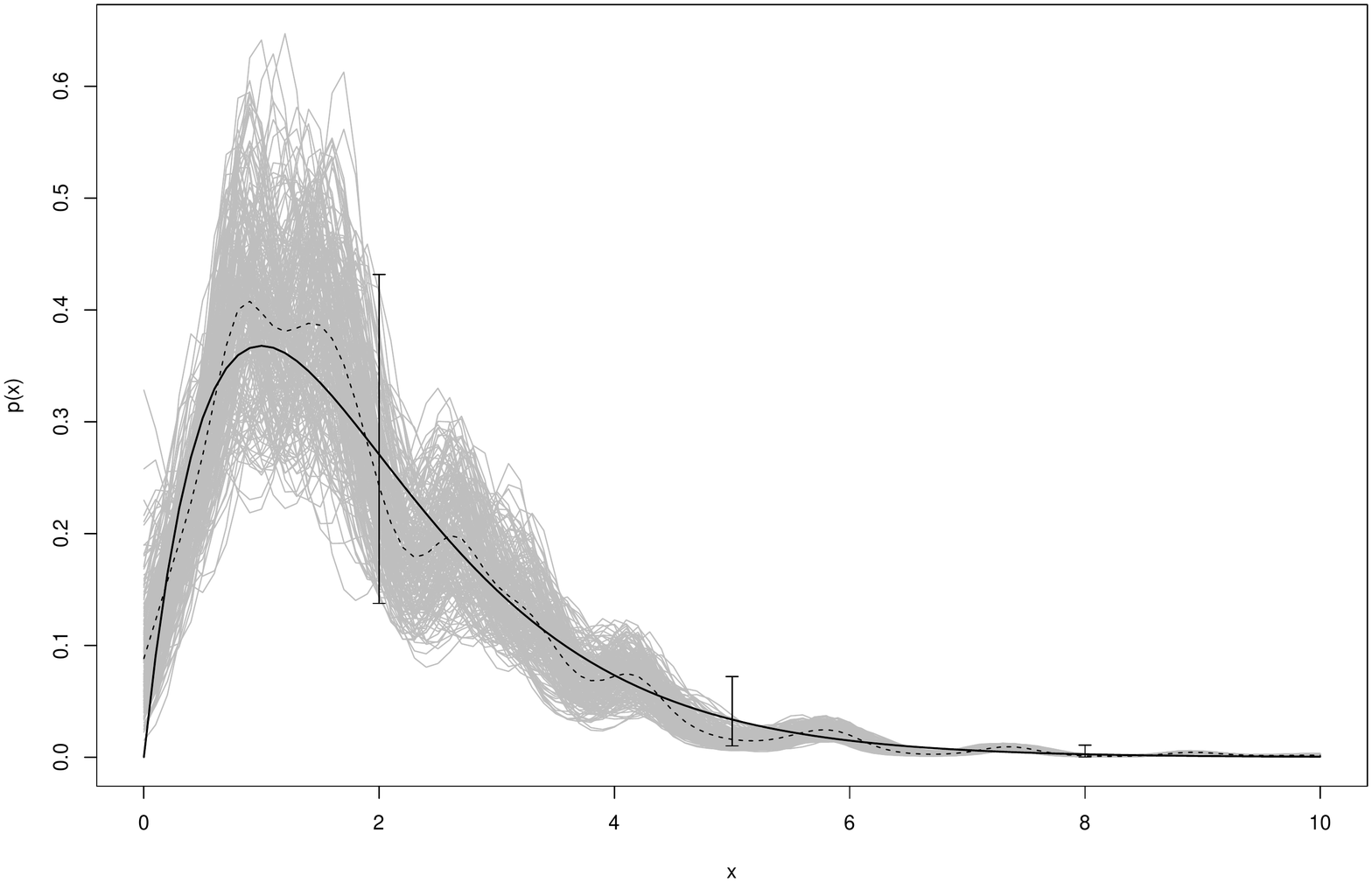}} 
\subfigure[Example 3-3]{\includegraphics[width = 0.3\linewidth]{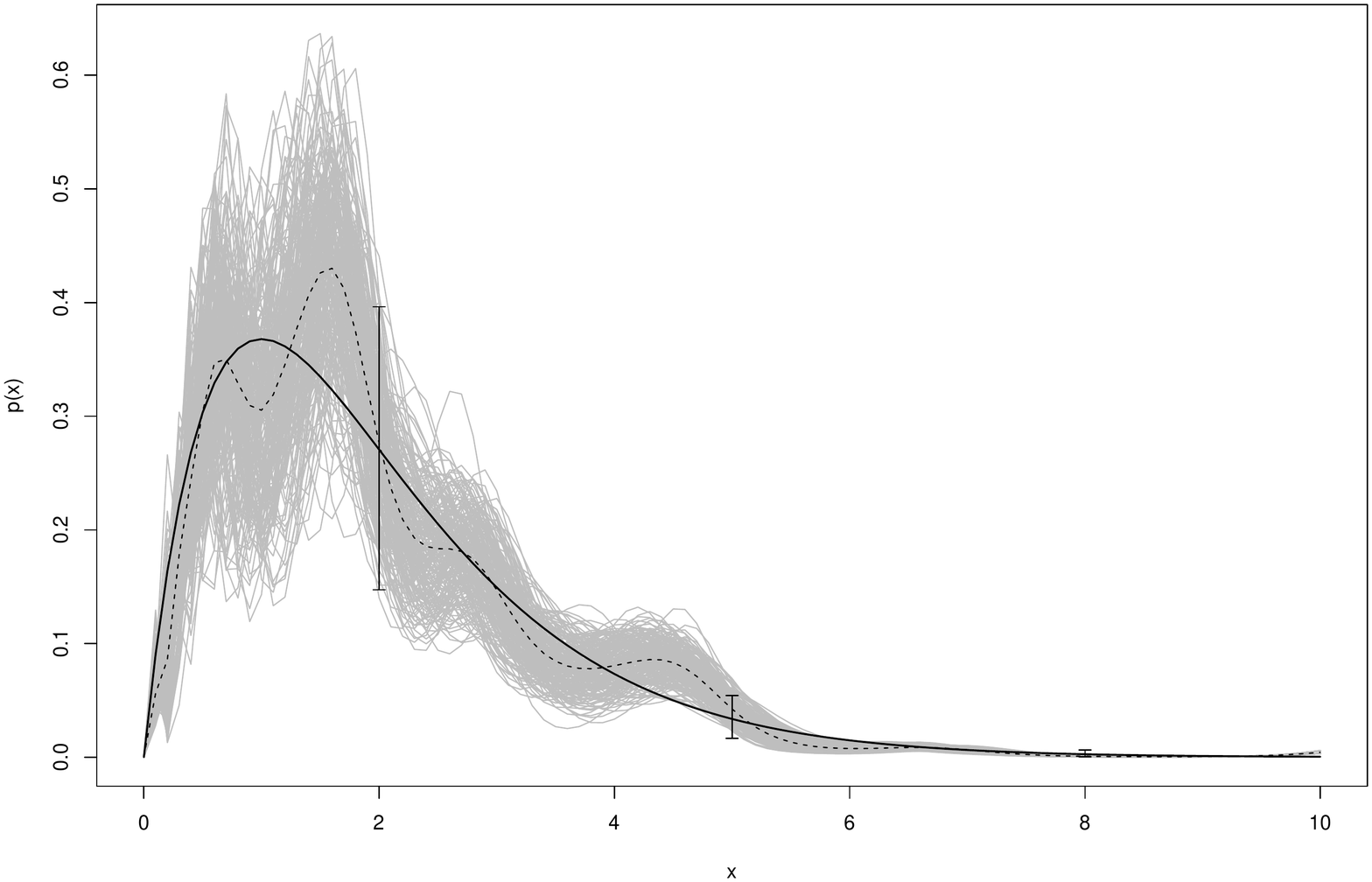}}
\end{center}
\caption{Plots of the predictive recursion mixing density estimates (gray) based on 200 random permutations of the data sequence, with the true mixing density (black) overlaid.  Dashed line corresponds to the predictive recursion estimate averaged over permutations. Vertical lines correspond to the central 95\% interval from the sampling distribution of $p_n(x)$, for $x \in \{2,5,8\}$.}
\label{fig:density}
\end{figure}

As a first visualization, we simulate $n=500$ observations from each of the above mixture models.  For each data set, we run the predictive recursion algorithm for 200 randomly sampled permutations.  Each panel in Figure~\ref{fig:density} shows these 200 density estimates, $p_n^S$, the true density, $p$.  As we expect, the cluster of permutation-based densities hugs the true mixing density rather closely throughout the range, with more variability in regions where the true density has more curvature.  Also displayed in these panels is a central 95\% interval from the sampling distribution of $p_n(x)$, at $x \in \{2,5,8\}$, based on 500 samples of size $n=500$ from the mixture model.  As suggested by \eqref{eq:unbiased}, the spread of the permutation distribution matches that of the sampling distribution relatively accurately at all three $x$ values and across all 9 of the examples.  

To assess our claim that the permutation-based uncertainty quantification is approximately valid, we repeat the above experiment 500 times, extract the nominal 95\% confidence interval for $p(x)$ based on the permutation distribution and check its coverage probability.  Table~\ref{tab:coverage} shows the estimated coverage probabilities for the all 9 examples, at each of the three $x$ values, and for two different sample sizes.  Note, first, that the coverage probability increases from $n=500$ to $n=1000$.  Major departures from the targeted 95\% level are at $x$ values around which $p$ has considerable curvature; in regions where $p$ is smoother, the coverage probability estimate tends to be closer to the 95\% level.  The overall message is that the permutation-based distribution gives approximately valid uncertainty quantification across a variety of mixing densities and kernels. 


\begin{table}[t]
  \begin{center}
    \begin{tabular}{ccccccc}
    \hline
    & \multicolumn{3}{c}{$n=500$} & \multicolumn{3}{c}{$n=1000$}\\
    \cline{2-7}
    Example & $x=2$ & $x=5$ & $x=8$ & $x=2$ & $x=5$ & $x=8$\\
    \hline 
    1-1 & 0.914 & 1.000 & 0.904 & 0.964 & 1.000 & 0.976\\
    2-1 & 0.882 & 0.990 & 0.882 & 0.956 & 1.000 & 0.952\\
    3-1 & 0.880 & 1.000 & 0.888 & 0.950 & 1.000 & 0.962\\
    1-2 & 0.994 & 0.476 & 0.948 & 1.000 & 0.488 & 0.982\\
    2-2 & 0.986 & 0.914 & 0.938 & 0.998 & 0.968 & 0.970\\
    3-2 & 0.972 & 0.910 & 0.918 & 0.998 & 0.966 & 0.972\\
    1-3 & 0.998 & 0.930 & 0.550 & 1.000 & 0.982 & 0.710\\
    2-3 & 0.994 & 0.862 & 0.378 & 1.000 & 0.938 & 0.540\\
    3-3 & 0.996 & 0.906 & 0.554 & 1.000 & 0.984 & 0.644\\
    \hline
    \end{tabular}
  \end{center}
\caption{Estimated coverage probabilities for the mixing density $p(x)$ in the nine examples across different sample sizes and $x$ values.}
    \label{tab:coverage}
\end{table}

\section{Conclusion}
\label{S:discuss}

This paper describes a simple permutation-based approach to uncertainty quantification about a mixing distribution by leveraging the built-in dependence of the predictive recursion estimator on the data ordering.  The development and numerical results presented here suggest that the uncertainty quantification achieved by this approach, e.g., a confidence interval for the mixing distribution or density function, are valid in the sense that the frequentist coverage probability is approximately equal to the interval's nominal level.  

We will end with two concluding remarks.  First, as noted in Section~\ref{S:examples}, we are currently lacking results on the pointwise consistency of the predictive recursion estimator of the mixing density estimate.  At present, we have only results on almost sure convergence of mixing measure estimator to the truth in the weak topology.  One idea is to prove the pointwise convergence directly using the structure of the predictive recursion estimator.  Another idea is to check the available sufficient conditions, namely, equicontinuity \citep[e.g.,][]{boos}, to convert weak convergence of measures into uniform convergence of densities.  Unfortunately, we were unable to push through either of these approaches, but this does not shake our confidence in the convergence conjecture.  

Second, the idea of leveraging data ordering dependence employed herein is not specific to the predictive recursion estimator.  That is, when data are independent and identically distributed or, more generally, exchangeable, and the density estimator in consideration depends on the data ordering, then the total-variance argument in \eqref{eq:unbiased} could be applied and approximately valid uncertainty quantification could be achieved.  A natural question is: are there any density estimators that depend on the data ordering?  Interestingly, while off-the-shelf estimators tend to be permutation invariant, one can consider Ces\'aro averages of these order-independent estimators, which retain the estimator's good asymptotic properties while simultaneously creating order-dependence that can be leveraged using the techniques presented here for valid uncertainty quantification.

\section*{Acknowledgments}

This work is partially supported by the U.S.~National Science Foundation, under grants DMS--1737929 and DMS--1811802.

\bibliographystyle{apalike}
\ifthenelse{1=0}{
\bibliography{/Users/rgmarti3/Dropbox/Research/mybib}
}

{

}

\end{document}